\documentstyle[wstwocl,epsfig]{article}
\def\np#1#2#3   {{\em Nucl.\ Phys.}\ {\bf#1} (#2) #3}
\def\pl#1#2#3   {{\em Phys.\ Lett.}\ {\bf#1} (#2) #3}
\def\prev#1#2#3 {{\em Phys.\ Rev.}\ {\bf#1} (#2) #3}
\def\prl#1#2#3  {{\em Phys.\ Rev.\ Lett.}\ {\bf#1} (#2) #3}

\newif\ifpreprint
\preprinttrue

\begin{document}


\title{NEW RESULTS FROM MIXED-ACTION LATTICE GAUGE THEORY}

\ifpreprint
\firstauthors{\hbox to \hsize{\rlap{\rm SWAT/87}\hfill Peter
Stephenson\hfill\llap{\rm hep-lat/9509070}}}
\else
\firstauthors{Peter Stephenson}
\fi

\firstaddress{Department of Physics, University of Wales Swansea, \\
Singleton Park, Swansea, SA2 8PP, U.K.\footnote{}}

\twocolumn[\maketitle\abstracts{
Lattice gauge theory involves a particular choice of discretisation of
the gauge action, notably in the representation of the gauge group.
Physical results should be independent of that
choice (an example of universality).  We use a combination of two
different representations: in our case, only one of these is irreducible.
We extend and attempt to clarify recent
results which showed possible problems for universality for the gauge
group SU(2).  We suggest that the presence of artifacts actually
distorts the mixed-coupling plane.  This implies that a separation of
artifacts from physical quantities is intrinsically difficult.  We
suggest what the nature of the resolution might be.  Full results will
appear elsewhere.
}]


\section{Introduction}
These days, lattice gauge theory is able to provide quantitative
results for an increasing number of problems in field theory.  It has
taken some two decades of trial, error and insight to reach this
position.  Nonetheless, it is still vital to test the formalism for
any possible problems lurking just around the corner from the results
presented as QCD phenomenology: in fact, there is less excuse than
ever for ignorance.\footnotetext[0]{${}^a$Address from $1^{\rm st}$
October 1995: DESY-IfH Zeuthen, 15735 Zeuthen, Germany.
\ifpreprint Talk given at EPS-HEP, Brussels, July 1995. \fi}

Many such tests probe the area of {\it universality}:  the choice of a
lattice-regularised theory is far from unique, but the same physics
should result in every case when one goes to the continuum limit and
hence the details of behaviour at the cut-off scale are expected to become
irrelevant. In this case, we investigate the dependence on the
representation of the gauge group used in the standard Wilson lattice
action:
\begin{equation}
S = \beta\sum_{\rm plaq}\mathop{\rm Tr}\nolimits_{\rm rep}
(U_{\rm plaq}).
\end{equation}

Here, $U_{\rm plaq}$ is an element of the gauge group
corresponding to the smallest closed path on the lattice, the
plaquette; $\beta$ is the inverse coupling for the representation
chosen.  The representation itself is hidden within the trace ${\rm
Tr}_{\rm rep}$.  If we choose our $SU(2)$ group elements to be
two-by-two unitary complex matrices, which we can write as four real
numbers with one constraint, the fundamental (spin-$1/2$)
representation of the gauge group just corresponds to the trace of the
matrix up to a factor $1/2$.  This is the most natural choice, and the
one that is usually made in lattice gauge theory.

The adjoint or spin-1 representation, however, contains the trace of
the matrix squared.  This means that the action is insensitive to a
change of sign in the matrices.  Pictured in terms of the SU(2) group
manifold, a three-sphere, the fundamental representation treats all
points on the manifold as distinct, while the adjoint representation
identifies the opposite ends of diameters of the sphere and
corresponds to the group SO(3) (as do all whole-integer
representations of SU(2)).

The continuum limit here is where the lattice spacing $a$ goes to zero
in physical units.  Because of asymptotic freedom, this occurs as
the inverse coupling $\beta$ goes to infinity, which is the
perturbative limit for the unrenormalised degrees of freedom (though
not for the physical ones).  Universality implies that the
topology of the gauge manifold should be irrelevant in this limit.

In the early eighties, Bhanot and Creutz\cite{bc}\ realised that one
could gain even more information by combining together terms in the
action with a trace in both representations,
\begin{equation}
S = \sum_{\rm plaq}\left(\beta_F\mathop{\rm Tr}\nolimits_{\rm
F}(U_{\rm plaq})
+\beta_A\mathop{\rm Tr}\nolimits_{\rm A}(U_{\rm plaq})\right),
\end{equation}
where $F$ and $A$ stand for the fundamental and adjoint
representations.  They then investigated the phase diagram in the
mixed-coupling plane $(\beta_F,\beta_A)$.

The resulting diagram was dominated by two first-order bulk
transitions, which joined to form another transition that ended
abruptly (see top left of figure~\ref{fig:piccie}).  As these are bulk
(volume) transitions, there is no physical scale associated with them
and they remain at a fixed coupling $\beta$: they cannot be taken to
the continuum limit and are therefore artifacts of some sort.

The transition intersecting the $\beta_A$ axis was explained in terms
of monopoles by Caneschi, Halliday and Schwimmer\cite{hs,chs}: the
monopoles were present because of non-contractable loops in the gauge
manifold, namely semi-circumferences whose end points were in this
case identified due to the $SO(3)$ invariance.  It was further shown
that the high-$\beta_A$ phase, beyond the monopole transition, was the
same as the phase in the standard $\beta_A = 0$ theory which has no
bulk transitions.

In fact, one can think of the separate area at the top left of the
phase diagram as being the region where two `copies' of the underlying
gauge system with its continuum limit exist, near the identity $I$ and
its negative $-I$ in the gauge manifold (these are of course identical
in the pure-adjoint case), related by a Z(2) symmetry (i.e.\ the
regions are really the same gauge system reflected) which can be
broken by increasing $\beta_F$.  In the rest of the diagram, there is
only one `copy', near $I$.  As one goes towards the continuum limit of
small lattice spacing and large $\beta$, only the gauge fields lying
near $I$ and (in the one case) the image near $-I$ are important.
Consequently, universality is not in danger; the lines due to the
extra Z(2) degree of freedom are merely an annoyance.


\section{Finite temperature effects}
This complicated but comprehensible picture was spoilt recently when
Gavai, Grady and Mathur\cite{mgg}\ investigated the finite temperature
transition known to exist in fundamental ($\beta_A=0$) SU(2).  This is
different in nature to the bulk transitions described above: it
depends on a physical scale, here the critical temperature $T_c$.  In
the usual formalism on a lattice of physical size $(N_sa)^3\times N_ta
= L_s^3\times L_t$, where $N_s$ and $N_t$ are integers, $T_c = 1/L_t$.
Increasing $N_t$, one therefore needs to decrease $a$ and hence
increase $\beta$ to recover $T_c$, which leads one again to the
continuum limit.  In other words, as one increases the number of links
$N_t$ in the time dimension of the lattice, the transition should move
to larger $\beta$.  This has been confirmed and quantitatively
analysed for the usual fundamental case\cite{efmwhk}.

Universality requires that this transition extend into the mixed
coupling plane: in fact the most na\"\i ve picture in the perturbative
limit would be an ellipse extending from the $\beta_F$ to the
$\beta_A$ axis.  In the continuum limit it should also clearly
separate itself from the bulk effects, which have no such limit.

However, refs.\cite{mgg}\ show that, on the contrary, the transition,
which is second order for $\beta_A=0$, becomes first order and
furthermore seems to merge with the line with the endpoint seen by
Bhanot and Creutz (which appeared to retain a finite temperature
nature), in an apparently clear violation of universality.  This is
the problem we try to elucidate here.


\section{Simulations}
We have actually simulated using the SO(3)-invariant action proposed by
Halliday and Schwimmer\cite{hs}.  Instead of a pure adjoint part of
the action they added an auxiliary Z(2)-valued field $\sigma$, defined
on the plaquettes:
\begin{equation}
S = \sum_{\rm plaq}\big(\beta_F\mathop{\rm Tr}\nolimits_{\rm
F}(U_{\rm plaq})
+\beta_V\mathop{\rm Tr}\nolimits_{\rm
F}(U_{\rm plaq})\times\sigma_{\rm plaq}\big),
\end{equation}
where the path integral measure is extended to include a sum over all
values of $\sigma = \pm1$.  They called this the Villain form of the
action.

The $\beta_V$ term does not correspond to an irreducible
representation of SU(2) and indeed when decomposed includes all
whole-integer representations (in the language of spin) of the group.
However, the SO(3) nature is presumably the dominant effect at work
and indeed the phase diagram found for the $(\beta_F,\beta_V)$
plane\cite{chs} is very similar to that of Bhanot and Creutz,
although it should be noted that the scale on the SO(3)-invariant axis
is somewhat different.

The main advantage of the Villain action in this work is technical:
its form ($\mathop{\rm Tr}\nolimits_{\rm F}$ is linear in the SU(2)
matrix algebra used while $\mathop{\rm Tr}\nolimits_{\rm A}$ is not)
means an efficient heatbath-plus-overrelaxation updating scheme can be
used for the Monte Carlo analysis in both the gauge and Z(2) degrees
of freedom rather than the less efficient Metropolis scheme required
in the adjoint case.


\section{Results}
\begin{table}
\caption{Critical couplings and orders for $N_t=2$ and 4.}
\begin{tabular}{r|ll|ll} \hline\hline
$\beta_F$ & \multicolumn{2}{c|}{$N_t=2$} & \multicolumn{2}{c}{$N_t=4$} \\
 & \multicolumn{1}{c}{$8^3$} & \multicolumn{1}{c|}{$16^3$} &
\multicolumn{1}{c}{$8^3$} & \multicolumn{1}{c}{$16^3$} \\ \hline
1.0 & 1.659(7) & 1.651(2) &  1.98(1)  & 1.983(4) \\
1.7 & 1.366(5)  & 1.359(3)  &  1.571(2)  & \\
2.2 & 1.1657(11)  & 1.1681(8)  &  1.2851(6)  & \\
2.5 & 1.0525(4)  & 1.0528(2)  & 1.13(1)  & \\
3.0 & 0.8659(1)\rlap{ $(6^3)$}  & & 0.87--0.93  & \\
\hline\hline
\end{tabular}
\label{tab:results}
\end{table}

We have found the phase transition in $\beta_F$ for a range of
different $\beta_V$.  We have used both $N_t=2$ and $N_t=4$; on the
latter, our results so far only consist of one spatial volume with
$N_s=8$, so extracting the order of the phase transition is less
precise than on the $N_t=2$ lattices, where we have $N_s=6$ (for
exploratory work) $N_s=8$ and $N_s=16$.  Our simulations typically
consist of 80,000 sweeps, each consisting of one heatbath plus four
over-relaxation steps, so that we expect autocorrelations to be low
(although they are taken into account in the bootstrap error
analysis).

The results, obtained from the peak in the susceptibility of the
Polyakov loop, are shown in table~\ref{tab:results}.  As for the order
of the phase transition, in the $N_t=2$ case it appears to be second
order up to $\beta_V=2.5$ and in the $N_t=4$ case up to $\beta_V=2.2$,
and first order thereafter.  In these intermediate cases, this is not
easy to determine definitely and we quote the best estimate.  We shall
require more simulations right on the phase transition for this to be
clearer.  In the other cases we can be more definite, as discussed
below.

\setlength{\unitlength}{0.7mm}
\begin{figure}[hbt]
\begin{picture}(100,110)(0,1)
\mbox{\epsfxsize8.0cm\epsffile{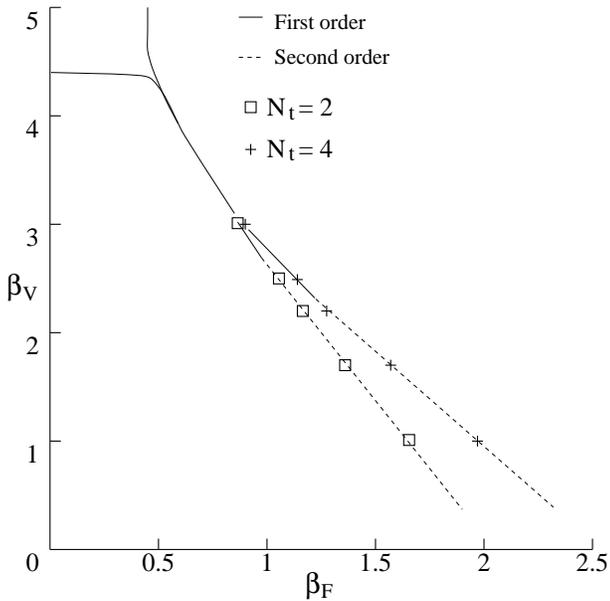}}
\end{picture}
\caption{The fundamental/Villain plane: see text for description.}
\label{fig:piccie}
\end{figure}
The results are also shown in figure~\ref{fig:piccie}.  Only the
squares and crosses represent actual results; the lower lines are
drawn by hand as a guide.  The upper lines are also drawn by hand
after the results of ref.\cite{chs}.  Presumably the lines join
somewhere (see below).

The basic picture is the same as that of ref.\cite{mgg}, and the
trends seem clear despite the provisional nature of the results
presented.  At small SO(3) coupling $\beta_V$, the transition is
second order and the analysis is standard\cite{efmwhk}.  At large
$\beta_V$ the transition becomes first order, with a clear two-state
signal, and increasingly strong in the sense that it takes many Monte
Carlo sweeps before changing from one phase into another.  In fact on
the $8^3\times4$ lattice at $\beta_V=3.0$ we are only able to quote an
upper and a lower limit (although these are firm) for the phase
transition as between these values of $\beta_F$ both phases are stable
over several tens of thousands of sweeps.  Nonetheless, perhaps the
central observation of this paper is this: we do definitely observe an
$N_t$ dependence in this region, confirming that the transition is not
bulk even though it is first order.

Note, however, that even with $N_t=2$ we could establish the position
of the phase transition accurately only for $N_s=6$, which leaves the
possibility of finite spatial size effects, the more so when one takes
into account the convergence noted in the next paragraph.

One major difference from the previous results is that it is now much
clearer that the phase transition lines are roughly straight over a
wide range---itself a departure from na\"\i ve expectations of
universality---and that they converge towards the end point of the
bulk lines.  However, we have not verified the monopole and
Z(2)-symmetry-breaking bulk transitions in detail and the point of
convergence is unknown.  Indeed, as those transitions are also
strongly first order, it is very difficult to pin the transitions down
for lattices larger than those used in ref.\cite{chs}\ and finite size
effects in this region are impossible to rule out.


\section{Implications}
It now seems clear that the (presumably physical) finite size
transition somehow becomes entangled with artifacts, and that this
requires some sorting out.  A truly analytic understanding is some
way off; we make the following remarks merely in an effort to make the
first steps.

Under the assumption that universality of the gauge theory can be
salvaged, the second order finite temperature transition must actually
be present, but somehow hidden by the artifacts.  Therefore, we
suggest that the first order effects are distortions of the
mixed-coupling plane due to long-range effects of the bulk transition
lines.  If these effects were not present, the finite temperature
lines at increasing $N_t$ would simply move further out to large
$\beta$.  As they seem to be funnelled in towards the meeting point of
the two bulk lines, we suggest that this is covering up a whole range
of physics of different lattice spacings $a$: for some (hypothetical)
small $a$, the scale of $T_c$ takes one left of the point, and for
some large $a$, this scale emerges again and starts moving to larger
$\beta$.  Near the point, the distortions caused by this actually
crease the plane somehow, showing the first order effects we see which
hide the physical transition.

This is more natural than it may appear at first sight.  First of
all, the two bulk lines presumably have the effect of a fold---that
is, a range of physics is covered by a single value of $\beta_V$ or
$\beta_F$.  Their conjunction might then be expected to have this
effect in both directions, acting as the funnel observed.  Secondly,
long range remnants of the bulk effect certainly seem to be present
anyway.  There will be more discussion of this in the eventual paper.


\setcounter{secnumdepth}{0} 


\end{document}